# Qualitative and quantitative evaluation of a methodology for the Digital Twin creation of brownfield production systems

Dominik Braun
Graduate School of Excellence advanced Manufacturing Engineering (GSaME)
University of Stutgart
Stuttgart, Germany
dominik.braun@gsame.uni-stuttgart.de

Nasser Jazdi
Institute of Industrial Automation and Software Engineering
University of Stuttgart
Stuttgart, Germany
nasser.jazdi@ias.uni-stuttgart.de

Wolfgang Schlögl
Digital Engineering
Siemens AG
Nuremberg, Germany
schloegl.wolfgang@siemens.com

Michael Weyrich
Institute of Industrial Automation and Software Engineering
University of Stuttgart
Stuttgart, Germany
michael.weyrich@ias.uni-stuttgart.de

*Abstract*— The Digital Twin is a well-known concept of industry 4.0 and is the cyber part of a cyber-physical production system providing several benefits such as virtual commissioning or predictive maintenance. The existing production systems are lacking a Digital Twin which has to be created manually in a time-consuming and error-prone process. Therefore, methods to create digital models of existing production systems and their relations between them were developed. This paper presents the implementation of the methodology for the creation of multi-disciplinary relations and a quantitative and qualitative evaluation of the benefits of the methodology.

## I. Introduction

The increasing demand for consumer individual products in decreasing time is one of the current challenges for current production systems for discrete products [1]. The industry 4.0 initiative addresses these challenges with integrated communication between all involved components and adding a cyber part to the production systems to create cyber-physical systems (CPS) [2]. This enables supportive applications such as predictive maintenance or virtual commissioning and increases the flexibility of the production systems. To get there the physical production system needs to be extended with a cyber part representing the system and its capabilities. This refers to the Digital Twin of a production system as its cyber part [3]. The Digital Twin is an important concept and is discussed in several publications [4]. It is defined differently and encompasses other elements depending on the intended use case and the sector. This paper refers to the definition of Ashtari *et al.* [5] since they conducted several publications and extracted the common and necessary elements of the Digital Twin. Based on their research the main parts are models from the mechanic, electric, and software domains as well as the intra- and inter-disciplinary relations between these models. They outline the digital replica of a system (see Fig. 1).

Besides that, three characteristics have to be possessed to be a Digital Twin: First, there must be simulatable models but not only, to cover the dynamic behavior of the system; Second, the static models need to be synchronized with the physical system to represent the current state of the system; and third, an active data acquisition is needed to mirror the current dynamic activities of the system.

Recently developed production systems are meanwhile often developed according to the model-based system engineering approach and therefore already have most of the digital models available. Therefore, they can benefit from virtual commissioning [6] or reconfiguration support [7] using a Digital Twin with little effort. However, older production systems usually have no documentation or appropriate digital modeling. This is a problem because the digital twin must then be created for many existing production systems in a manual, time-consuming, and error-prone reverse engineering process in order for them to be able to overcome the new challenges in the future. Structure: The next section briefly conducts the state of the art regarding the approaches to creating the digital replica of existing production systems. Afterward, section III presents a methodology to create the digital replica as the base for the Digital Twin with a focus on structure and relations. Section IV describes the implementation of the methodology which will be used for the following evaluation. The following section V contains the evaluation description, results, and discussion. The last section VI finally summarizes the withdraws and gives an outlook.

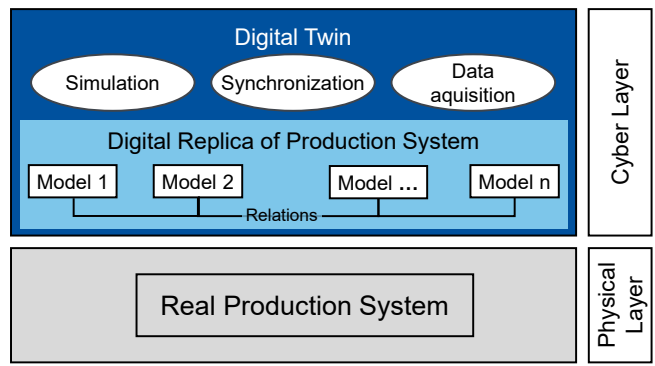

Fig. 1.  Digital Twin architecture according to Ashtari et al. [5]

## II. STATE OF THE ART

This section briefly concludes the current state of the art of the creation of models and their relations as the base of the Digital Twin of automated production systems, especially in the manufacturing domain. During the engineering of new production systems model-based system engineering is often used. This development approach uses various models to describe and exchange information all over the engineering domains and phases. This simplifies the reuse and exchange of information in contrast to document-based system engineering. Furthermore, model-based system engineering has a positive side effect regarding the Digital Twin of such systems, because these models are the necessary base of the Digital Twin [8]. They form, together with the relations between the models, the digital replica.

Existing production systems are lacking from these models because they are engineered some time ago when model-based systems engineering was not common yet. The mostly paper-based documentation of the production system is often missing or outdated due to poorly documented previous system modifications [9]. For this reason, reverse engineering of the production system is necessary before reconfiguration can be planned, either with or without a Digital Twin. To avoid errors, reduce manual effort, and enable the benefits of a Digital Twin for existing production systems, an automated method to digitalize these systems is necessary.

The most mentioned method in literature is in this context the mechanical CAD model creation using laser scanning [10, 11] or similar picture-based methods such as photogrammetry [12]. These optical methods use either laser scanners or modern cameras as a cheaper version to collect images of the production system [13]. Software is used to combine the images from several sources to calculate a 3D view and furthermore it optionally segments the 3D monolithic model into sub-segments [10]. Besides that, these methods do not create any other models from other domains or relations towards them.

Another method is the document analysis which consumes paper-based documentation. For this purpose, there must be up-to-date documentation of the information which should be modeled. This is most commonly the case with electrical circuit diagrams which are placed aside from the real manufacturing system and contain electrical changes in red color, so-called redlining. The document analysis uses image processing and optical character recognition methods to analyze the lines, symbols, text, and connections to build the current models. This recognizes the circuit elements and the manual changes to rebuild the electrical model. Apart from the electrical model no other models or relations are created by this method with the redlined electrical circuit documents [14].

Furthermore, the communication network is used as an information source by different methods. One scans the network for the participants to generate the list of existing devices [15], and others analyze the data captured from the communication network to retrieve changes and adapt the simulation accordingly [16].

Besides that, there are some methods to generate new models from other existing models or to reuse and combine existing models. These methods build, besides modes, at least relations between two domains, the domain of the source and target models. This is done not by many other methods, especially not the domain-specific methods described earlier. There is only one methodology for creating inter-domain relations for the Digital Twin, which is called anchor point method [17]. This methodology uses the PLC code, applies a naming convention based on industry standards to the code, and analyses the naming of the tags, function blocks, and data blocks to extract the structural changes and relations between the domains.

Overall these methods are mostly designed to automate the creation of specific models belonging to one domain and do not re-engineer the relations between the domains and production system parts. Only the anchor point method creates or updates inter-domain relations but it is designed to work with the new production system. It cannot be applied to brownfield systems because there must be a naming convention applied to the PLC code and Digital Twin models during engineering. Furthermore, these methods are most commonly used in research but are not yet applied in the industry besides the optical methods to create CAD models and the paper-based circuit diagram analysis.

In the process industry, there is a methodology [18] that addresses the holistic creation of the Digital Twin, but it focuses on the specific features and models of continuous processes (e.g. P&ID) and is not applicable to discrete production systems.

For the creation of the Digital Twin of existing discrete production systems, a methodology for the reverse engineering and creation of the relations without an existing Digital Twin is still missing.

## III. METHODOLOGY

A methodology to support the automated creation of a Digital Twin of existing production systems is presented in [19]. Because there is no data source containing all information needed at once (compare section II), this methodology uses three data sources to gain knowledge about the production system (see Fig. 2). The first data source is the PLC code, which is analyzed based on a rule set to extract functional relations between mechatronic components (sensor and actuators) and software components [20]. The second data source is the IO data from the PLC collected during runtime and stored in a time-series database. The last source is the position data of the product captured with a real-time locating system (RTLS) and stored as a time series in a database [19].

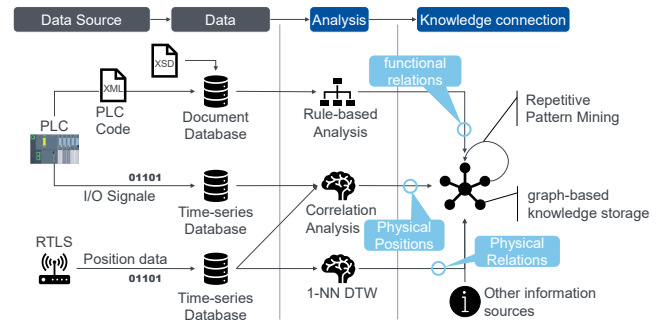

Fig. 2. Methodology – knowledge inside the graph-based storage contians the system relations and is used to build the digital replica

The PLC code is analyzed to reverse engineer the functional groups of the production system and their structure. This delivers the functional relations of the components and their related software function block from the PLC. The extracted information about the production system is mainly based on relations and forms a graph-like structure. For this reason, the information is stored in a graph database and the nodes and edges are provided with additional key-value pairs as labels, which specify them in more detail. To merge this information with the information from the other sources a hybrid hierarchy-based modeling approach [21] is applied to the graph structure and labels to make them machine-readable. This is done using the basic ontology of the approach in [21], which is based on standards on mechatronic systems and Digital Twins. The system-specific terminology box in this approach must be adapted to the system at hand or removed for other systems. This results in a less specific and more general description of the graph. The other information sources are analyzed to retrieve physical groups of components [19]. Therefore, a data-driven analysis of the position data from the RTLS and the IO data from the PLC both separately and in combination is used to estimate the physical structure, rough component positions, and component grouping. This allows us to group the components according to their physical relation [19]. Finally, the graph data is processed using frequent subgraph mining to identify often-used structural elements [22]. The identified, repetitive patterns can be summarized to reduce the graph size, which helps an engineer who has to work with this graph. Furthermore, this simplifies the reuse and modification of the digital models of commonly used assemblies. Overall, the methodology automates the creation of the relations of a digital replication and thus formed an important basis for the construction of a Digital Twin of an existing production system. The automated creation significantly reduces the required knowledge and time, and thus the dependency on expensive experts and the delays caused by human errors.

IV. IMPLEMENTATION

The methodology was prototypically implemented as a software assistance system. The software architecture is visualized in Fig. 3. There are two modules to translate and export data into standardized exchange formats. One is the export application which opens a native PLC Project in Total Integrated Automation (TIA) Portal through the TIA Openness API and converts it into XML files. The second is the Automation Markup Language (AML) export which process relation-based information from the neo4j database and exports it based on the node types into an AML file.

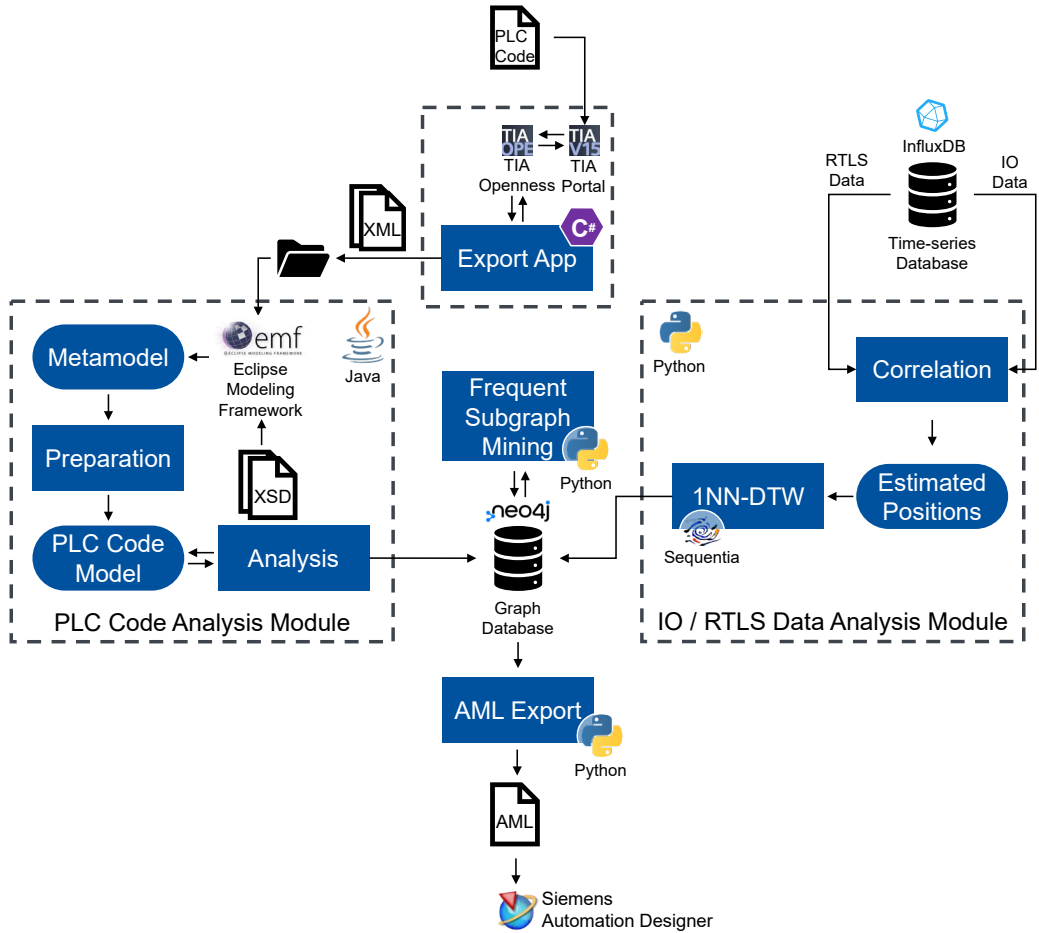

Fig. 3. The architecture of the software assistance system

Besides that, there are two software modules to analyze the data sources to gain information about the production system and its functional and physical relations and store them in the graph database.

The PLC code analysis module consumes the PLC code to retrieve the functional structure and grouping from it by a rule-based analysis [20]. It starts the vendor-specific export application to convert the PLC code and loads the PLC code in standardized XML format from the defined exchange folder. This implementation uses the Siemens TIA Openness API to translate the PLC code from TIA Portal v15 to XML. The XML files are then imported as java classes using the eclipse modeling framework into the metamodel. The preparation submodule extracts the core information from the vast number of classes and adds attributes to them to explicitly note the implicit connections between the classes to simplify the analysis process. This results in the PLC code model, which is finally analyzed by the Analysis submodule to separate the PLC code elements into functionally related groups. These elements, their attributes, and especially the grouping are stored in a neo4j graph database.

The second analysis module implements the data-driven analysis of the RTLS and IO data. Therefore, the physical grouping and the approximate position of the components are determined by analyzing the time correlation between moving material and changing IO signal and calculating the average position [19]. Therefore, the IO data is recorded from the OPC interface of the PLC and stored in an influx time-series database. Besides, a 1-nearest neighbor algorithm with dynamic time warping is trained with a labeled position data set to solve the multivariate time series multi-class classification problem. The trained model is used to assign the system components (sensors, actuators) to the location group classes based on the estimated position. Instead of classification, clustering can be used as well, which reduces the needed effort for the dataset creation and training but may result in a different split with reduced accuracy of the grouping. This is because the position data is not distributed as spatial groups, but are distributed along continuous trajectories and flow seamlessly from one cluster to the next.

The single module for frequent subgraph mining uses an implementation of a graph-based substructure pattern mining (gspan) algorithm to identify repetitive node structures and mark them as an instance of a specific template.

### A. Preparation for operation

To run the implemented methodology first data from the existing production system is needed. Since the advance of data-driven algorithms (machine learning) and the consequent growing importance of data, many system operators are already collecting IO data. The increasing usage of mobile robots (AGVs or AMRs) boosts the application of indoor localization systems and the collection of material positions, e.g. for documentation purposes. This data is needed for the analysis and can be used if already available or need to be collected. For the used evaluation system (described in section V.A) the data was collected in three hours using an OPC UA client collecting data from the PLC and storing it in an influx DB. The position data is collected using the RTLS and a python script tapping the data and assigning labels to the data based on user input.

Because the PLC code of the evaluation system was implemented with another TIA Portal version, the metamodel of the PLC code analysis needs to be updated. Therefore, the PLC code was exported as an XML file using TIA Openness and its scheme definition was extracted using the tool LiquidXML studio. The scheme definition was used to update the metamodel through the eclipse modeling framework EMF. This process took another four hours. After these preparations, the implemented methodology is ready to be applied to the evaluation system and reengineer the relations of the Digital Twin. If the assistance system is to systems with the same PLC vendor and the same version number, the metamodel update step can be skipped. For other vendors or versions, the update needs to be done once for the first application.

### B. Application to the brownfield system

First, a neo4j graph database is started to store the analysis results. Then the described software assistance system (see Fig. 4) is started where the analysis applications can be triggered via buttons to analyse the PLC code, the I/O, and RTLS data. This results in a graph containing the relational information about the system. The application contains a button that visualizes the most important and expressive relations that a human can verify the results. The results can be further examined by running cypher queries in the graphical user interface of the graph database. Frequent subgraph mining enriches the results by indicating repetitive node-relation structures starting from the system's root node. The automation devices such as the PLC or IO devices are excluded during the pattern mining otherwise e.g. an input device with its according eight sensors is found as main pattern. After mining and marking the structural pattern of function and physical groups the connection to the automation devices can be found in the graph database.

### C. Export to the modeling software

The analysis tool finally exports the retrieved and identified relations and repetitive structures as an AML file, which can be imported into a modeling software for Digital Twins. Siemens Automation Designer is the target software to evaluate the re-engineered system relations as part of the Digital Twin. For other software tools, either they can import AML as well or the export module needs to be replaced with a specific export function.

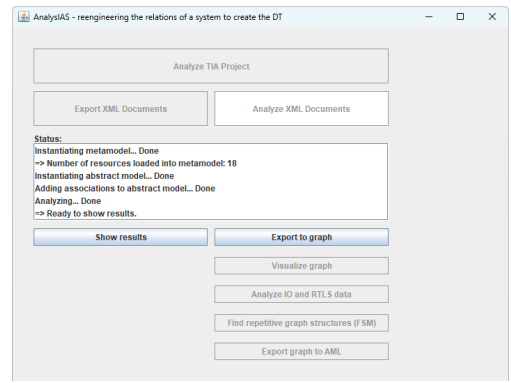

Fig. 4.  GUI of the software assistance system

## V. EVALUATION

This section describes how the methodology and its benefits are evaluated against the current manual Digital Twin creation process. The first section V.A describes the evaluation approach, and subsequently, section V.B presents and discusses the evaluation results which are finally concluded in the last subsection.

### A. Evaluation Approach

The methodology is evaluated by comparing the manual process of the Digital Twin creation for an intelligent warehouse with the automated process using the methodology described in section III. The intelligent warehouse is a part of the flexible production system (FlexCell) project in the ARENA2036 (Active Research Environment for the Next generation of Automobiles 2036) at the University of Stuttgart. This production system is a small research and demonstration system representing industrial systems and was built accordingly together with the three industry project partners Siemens, Kuka and Trumpf. The system consists of the intelligent warehouse, an autonomous mobile robot (AMR), and a laser welding machine (see Fig. 5). Each sub-system has its own PLC and acts independently. Besides these real systems, their Digital Twins were built manually during the project to demonstrate a closed software toolchain during engineering. The intelligent warehouse consists of two working levels for providing and withdrawing trays to the robot. Each level consists of four storage rows with several separate storage places. The intelligent warehouse catalogs, identifies, and provides four different raw metal sheets separated in trays. Therefore, it is equipped with 35 sensors and 25 actuators as well as its own PLC for control purposes.

The intelligent warehouse is used to evaluate the described methodology in section III. Therefore, the digitalization of this intelligent warehouse in Siemens Automation Designer is done once manually by the authors and some of the project partners. Besides that, the manual effort for the manual creation was conducted during ten interviews with more experts from academia and industrial companies such as Siemens, Kuka, ABB, or Lapp. All these experts know the intelligent warehouse from the project, or corporations, or got a short presentation and explanation of the warehouse and have some expertise with the Automation Designer software. The mean duration of the manual tasks is then compared with the automated digitalization of this system using the methodology as described in section IV.

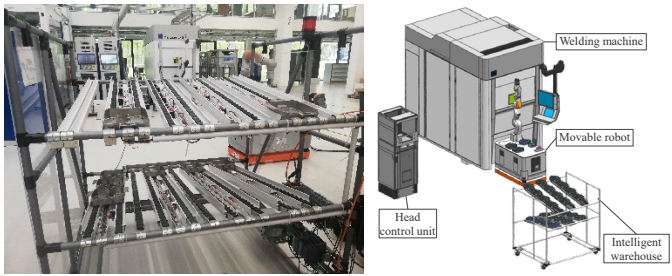

Fig. 5. FlexCell System in the ARENA2036 with the intelligent warehouse as picture (left) and CAD model (right)

### B. Evaluation Results and Discussion

This section presents the gathered time needed for the manual process and the duration of applying the presented methodology. Fig. 8 illustrates the comparison of the duration of the manual (left bar) and automated Digital Twin creation (right bar).

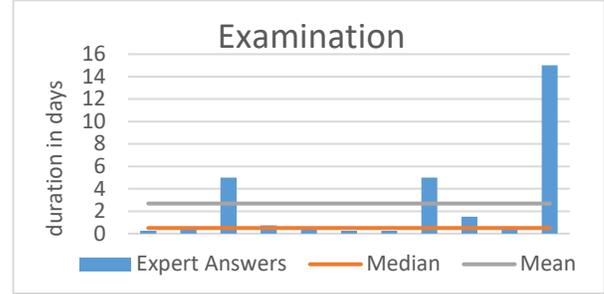

Fig. 6. Difference between mean and median due to outlier in the first question

To calculate the duration of the manual process a week was defined as five days with eight hours. The overall duration calculated as the median is seven days as in Fig. 8. The duration calculated as the mean is nine and a half days. The difference results from one answer regarding the (pre-) examination of the intelligent warehouse which was an outlier and causes a bit more than two days difference between mean and median (see Fig. 6). The other questions do not show effects through outlier and mean and median are nearly the same (see Fig. 7). Therefore, the duration of each step was calculated as the median because it is less affected by outliers and reduces the effect of missing expertise in single steps of this one expert.

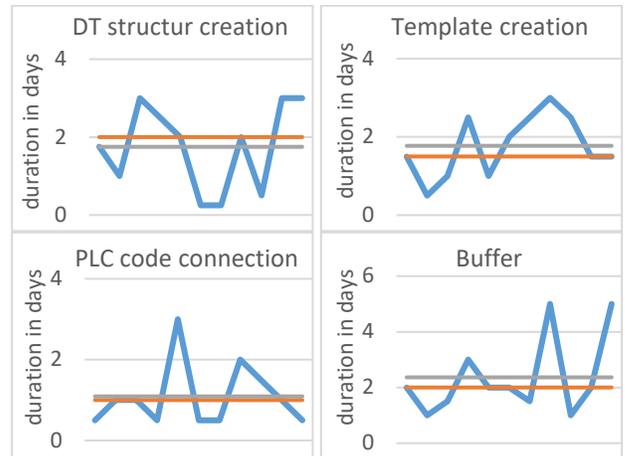

Fig. 7. Comparison of the mean (gray), and median (orange) in contrast to the answers of the experts (blue) for questions two to five

The process of automated creation represented as the right bar (see Fig. 8) was carried out with the implementation and process described in section IV on the intelligent warehouse. The automated creation of the relations as the base of the Digital Twin took around two days. The run time of the implemented software assistance system to analyze the data and create the relations was less than half an hour. The data collection, preparation, and the metamodel extension delivered the major part.

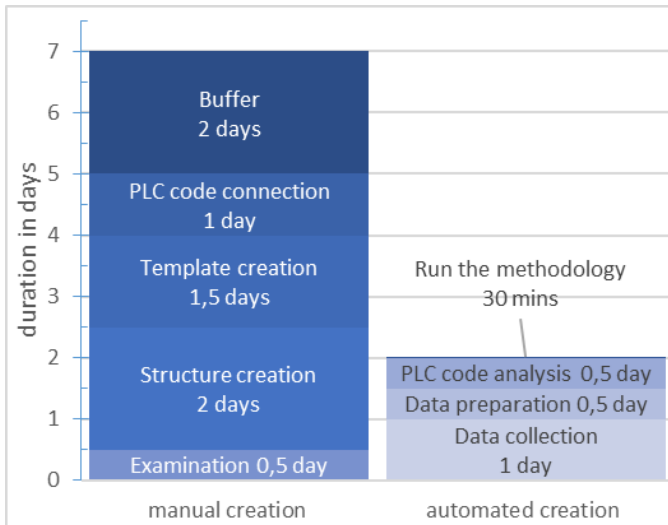

Fig. 8. Comparison of the duration of the manual and automated process of creating the digital twin of the intelligent warehouse

The manual process starts with an investigation process of the technical system, in this evaluation of the intelligent warehouse system. Most experts scheduled around two hours up to approximately one day. Only one answer is outside the common answer range of 15 days. This outlier is discharged using the median for the average duration calculation of half a day. Most experts would inspect the software, sensors, and actuators. As often as the previous the experts would inspect the mechanic but find this information a bit less important. Only two-thirds of the experts would examine the wiring between the components and the PLC, and less than half analyze the fieldbus connection and only with the lowest priority. The examination of the intelligent warehouse is followed by the step of creating the system structure in Automation Designer. This is with two days the largest part of the manual process and covered by the analysis part of the automated methodology. The following template creation step to identify and instantiate repetitive structures is estimated with one and a half days. The automated template identification of the methodology is supporting this step by identifying the repetitive structures in this system. The experts raised the argument this step can take more time if not only templates for this system are created but as well general templates for other systems. Templates are automatically identified with the methodology and the structure created but the templates are only generally usable if the analyzed system is generic for other systems or data of multiple representative systems is needed as input of the automated methodology. Besides that, there are no automatic naming rules created during the template creation. Thus, manual work is still needed to automate the component naming for reuse in other projects. The last step is to connect the PLC code and link the function- and data block as well as to connect the signals to the PLC channels. The experts assess the difficulty of this step rather low and estimate one day to complete it. Finally, the experts were asked to define a buffer time to cover their uncertainty about the duration of the entire manual process. The average buffer time is two days.

### C. Summary and Assessment of the results

In total, the overall manual process is five days and 2 days of buffer time. The automated process takes half an hour but around two days of preparation, a total of approximately two days. Based on these results, it is evident that the methodology saves up to 70% with buffer time and 60% if the buffer is not needed of the time compared to the manual creation of the Digital Twin basic relations of an existing production system. Besides that, the methodology reduces faults happening during the manual creation of such structures. The experts indicate that the steps can be done once even faster but searching and removing structural faults due to the number of nodes and relations add extends the steps. The same applies to template identification because the template borders and structure may seem reasonable at first but during usage, it turns out that it is not. Such human mistakes due to a large amount of information about big systems and oversights in the creation lead to errors and increased creation time. For other even larger systems, this problem may even increase. The creation of the Digital Twin relations with the software assistance system as the implementation of the presented methodology is always creating the same result also for larger systems.

## VI. Conclusion and Outlook

The developed methodology addresses the basic steps regarding the creation of the relation of a Digital Twin of existing production systems. The paper presented the prototypical implementation of the software assistance system and the quantitative and qualitative evaluation of the methodology. Therefore, the automated creation of the relations using the software assistance system was run on the existing intelligent warehouse of the FlexCell system in the ARENA2036 to create the relations. The durations collected by this automated process are compared with the duration of the manual creation process. The time needed for the manual process is collected by a manual implementation process done by the authors and the estimation from ten experts. The evaluation depicts a time reduction of 60-70% depending on the buffer time needed during the manual process. Due to the current development of collecting data also for existing systems in order to be able to implement AI-based applications, the process of collecting and processing data may be shortened further. Besides that, the methodology has qualitative benefits: Human errors in arranging or overlooking components in a large structure are avoided and the knowledge required respectively the availability of experts from several disciplines is reduced.

The resulting Digital Twin structure is not a ready-to-use implemented Digital Twin. The methodology and the interviews are not covering the creation of detailed models such as CAD, ECAD, or others besides the Digital Twin structure. To use the Digital Twin for a dedicated application e.g. virtual commissioning, models in the needed modeling depth have to be created or instantiated in the created Digital Twin structure if they already exist. This time is not covered by the evaluation neither the manual nor the automated process. The creation can be done manually or by other automated methods such as laser scans. The applied basic ontologies based on standards to the graph database enable an automated junction of information from other methods into the knowledge base for the Digital

Twin structure creation. Besides that, the methodology can be extended further with a mechanism to apply the naming rules of the anchor point method [17] during the Digital Twin creation and thus enable the synchronization of the Digital Twin in the future. To reduce the manual process for new PLC code vendors and versions, the metamodel of a commercial implementation can get the full XSD files containing all possible elements from the vendors and implement a metamodel based on them. This increases the applicability and decreases the duration of the automated process even further.

ACKNOWLEDGMENT

This work was supported by the German Research Foundation (DFG) within the Excellence Initiative – GSC 262 and the Ministry of Science, Research and the Arts of the State of Baden-Wurttemberg within the sustainability support of the projects of the Excellence Initiative II.